\documentclass[twocolumn,aps,prl,showpacs,superscriptaddress]{revtex4}

\usepackage{graphicx}
\usepackage{amsmath,amssymb}
\usepackage{color}

\begin{document}


\title{Hidden variable models for quantum mechanics can have local
  parts}


\author{Jan-{\AA}ke Larsson}
 \email{jalar@mai.liu.se}
 \affiliation{Institutionen f\"or Systemteknik och Matematiska
   Institutionen, Link\"opings Universitet, SE-581 83 Link\"oping,
   Sweden}

\author{Ad\'an Cabello}
 \email{adan@us.es}
 \affiliation{Departamento de F\'{\i}sica Aplicada II, Universidad de
 Sevilla, E-41012 Sevilla, Spain}


\date{\today}



\begin{abstract}
  We criticize Colbeck and Renner's (CR's) statement that ``any hidden
  variable model can only be compatible with quantum mechanics if its
  local part is trivial'' [Phys. Rev. Lett. {\bf 101}, 050403 (2008)].
  We note that CR's attempt to divide a nonlocal hidden variable model
  into a ``local part'' and a ``nonlocal part'' contains a restriction
  on the latter. This restriction implies that the division is really
  into a ``local part'' and a ``nonsignaling nonlocal part.''  CR's
  nonsignaling requirement on the ``nonlocal part'' cannot be
  physically motivated, since the hidden variables cannot be accessed
  by experimenters. Nor is it a natural mathematical generalization
  from the local hidden variable case, since it is simple to make a
  generalization without CR's requirement.
  We give an explicit nonlocal hidden variable model that, in the case
  the restriction is not enforced, contains nontrivial local hidden
  variables.
\end{abstract}



\maketitle


{\em Introduction.---}If quantum mechanics is correct but not complete
\cite{EPR35}, a fundamental question is which classes of more detailed
theories are compatible with it. Experimental violations of Bell
inequalities \cite{ADR82, WJSWZ98, RKMSIMW01, MMMOM08} suggest, and a
loophole-free violation would confirm (see \cite{RWVHKCZW09} for a
proposal), that local hidden variable (HV) models \cite{Bell64,
  CHSH69} are not an alternative. Loophole-free experimental
violations of noncontextual inequalities \cite{RKMSIMW01, MMMOM08,
  KZGKGCBR09} already prohibit noncontextual HV models
\cite{Specker60, Bell66, KS67, Cabello08}.

Recently, considerable experimental effort \cite{GPKBZAZ07, BLGKLS07,
  BBGKLLS08} has been devoted to testing the violation of an
inequality proposed by Leggett \cite{Leggett03} and a different
inequality proposed by Branciard {\em et al.}  \cite{BLGKLS07}, valid
for specific classes of nonlocal HV models. These experiments suggest
that these models are not an alternative either. Another example is an
early theoretic study \cite{EPR92}, on nonlocal HV models that consist
of a statistical mixture of a local and a nonlocal ensemble, in which
case the local ensemble must have probability zero.

More recently, Colbeck and Renner (CR) have made a much more profound
and far-reaching claim. According to CR, ``any hidden variable model
can only be compatible with quantum mechanics if its local part is
trivial'' \cite{CR08}. Several experiments in progress are currently
testing the violation of an inequality that CR propose.

The present paper studies CR's formal result in detail, and
focuses on the fact that CR's division into ``local part'' and
``nonlocal part'' contains a nonsignaling restriction on the
``nonlocal part.'' The proof that the ``local part'' is trivial
in \cite{CR08} only holds under this restriction on the
nonlocal part. We scrutinize the weaknesses of the physical and
mathematical motivation for this restriction.  We also study
the effects of the restriction in an explicit nonlocal HV
model; local HVs contained in the model are forced to belong to
the ``nonlocal part'' to prohibit signaling in the ``nonlocal
part.''  From the lack of physical motivation for the
nonsignaling requirement, and the fact that it is imposed on
the ``nonlocal part'' rather than the ``local part,'' we
conclude that it should not be used, and thus, that nonlocal HV
models can have nontrivial local parts.

We will use the scenario of Bell \cite{Bell64} and the notation
of CR's Letter \cite{CR08}.  Consider two particles which
travel to two far apart locations, Alice's and Bob's, where a
local measurement is made on each particle.  The setting $a$
($b$) pertains to the local measurement $A$ ($B$) at Alice's
(Bob's) location and the corresponding outcome is denoted $X$
($Y$).  The nonlocal HV $\Gamma$ can be divided into a ``local
part'' at the first site $U$, a ``local part'' at the second
site $V$, and a nonlocal part $W$.  The distribution of
$\Gamma$ is not specified in Ref.~\cite{CR08}, but it appears
\cite{CRpriv} that the requirement is
\begin{equation}
  \label{eq:1}
  \begin{split}
    P_{\Gamma|ab}(\gamma)&=P_{UVW|ab}(u,v,w)\\
    &=P_{W|abuv}(w)P_{UV}(u,v),
  \end{split}
\end{equation}
and in the case that $W$ is trivial, the reader will recognize
Eq.~(\ref{eq:1}) above as the usual requirement on the distribution of
local HVs.

The probability distribution of the outcomes $X$ and $Y$ is then
determined by the values of these variables and is denoted
$P_{X|abuvw}$ and $P_{Y|abuvw}$. Now, CR proceed to ``ignore'' the
nonlocal part $W$ and require that the HV model obeys (see Eq.~(2) in
Ref.~\cite{CR08})
\begin{equation}
  \begin{split}
    \label{eq:2}
    P_{X|abuv}&=P_{X|au},\\
    P_{Y|abuv}&=P_{Y|bv}.
  \end{split}
\end{equation}
The reader will again seem to recognize the usual requirement on a
local HV model; that the local outcome cannot depend on the remote
setting [usually there is only one local HV $\lambda=u=v$, but the
present formulation is equivalent in the case that $X=Y$ whenever
$a=b$, i.e., given existence of EPR elements of reality].  However, if
you do not ``ignore'' the nonlocal part $W$, you obtain
\begin{equation}
  \label{eq:3}
  \begin{split}
    P_{X|abuvw}&=P_{X|auw},\\
    P_{Y|abuvw}&=P_{Y|bvw}.
  \end{split}
\end{equation}
We would argue that this is the correct generalization of requirement
(\ref{eq:2}) to the nonlocal HV case: The local outcome may depend on
the HVs, and the local setting of the measurement apparatus, but not
the remote setting.  Evidently, if $W$ is trivial, the expressions
(\ref{eq:3}) reduce to that in (\ref{eq:2}). If $W$ is nontrivial,
averaging over it will remove dependence of $w$ from the expressions
(and this appears to be what CR do ``when ignoring'' $W$). Under
requirement~(\ref{eq:3}), we obtain
\begin{equation}
  \begin{split}
    \label{eq:4}
    P_{X|abuv}&=\sum_w P_{W|abuv}(w)P_{X|abuvw}
    \\&=\sum_w P_{W|abuv}(w)P_{X|auw}
    =P_{X|abu},\\
    P_{Y|abuv}&=\sum_w P_{W|abuv}(w)P_{Y|abuvw}
    \\&=\sum_w P_{W|abuv}(w)P_{Y|bvw}
    =P_{Y|abv}.
  \end{split}
\end{equation}
That is, even if (\ref{eq:3}) holds, the dependence of the
remote measurement setting remains, mediated through the
nonlocal HV $W$. This in itself is not strange, but means that
requiring (\ref{eq:2}) enforces \emph{an additional
restriction} on the HV model.  This is described by CR in
\cite{CR08} as follows:
\begin{quote}
  In particular, knowing the value of the local hidden variables
  would not permit signaling between Alice and Bob.
\end{quote}
We would argue that this in fact is an explicit extra requirement, not
on the ``local part'' $U$ and $V$, but on the ``nonlocal part'' $W$.
Comparing requirement (\ref{eq:2}) with the usual non-signaling
requirement
\begin{equation}
  \begin{split}
    \label{eq:5}
    P_{X|ab}&=P_{X|a},\\
    P_{Y|ab}&=P_{Y|b},
  \end{split}
\end{equation}
it is clear that (\ref{eq:2}) should be read: ``For each value of $u$
and $v$, the remaining HV model should be nonsignaling.''  We do not
agree with CR that this is necessary; but we do agree with CR
\cite{CR08} that ``\ldots signaling may be possible given knowledge of
the nonlocal variable $W$.''  In fact, we would go further and say
that signaling is \emph{only} possible through the nonlocal variable
$W$. Indeed, given the value of the nonlocal HV, no dependence on the
remote setting remains for the local outcome.  This is the requirement
that should be used, and it is exactly Eq.~(\ref{eq:3}) expressed in
words. Thus, Eq.~(\ref{eq:3}) is the natural requirement, and not the
requirement (\ref{eq:2}) [that the ``nonlocal part'' should be
nonsignaling given the value of the ``local part'' HVs].

There is another more direct implication from CR's restriction
(\ref{eq:2}): The nonlocal HV model must be nonsignaling. This is
simple to see, since the restriction~(\ref{eq:2}) directly implies,
for example
\begin{equation}
  \begin{split}
    \label{eq:6}
    P_{X|ab}&=\sum_{u,v} P_{UV}(u,v)P_{X|abuv}\\&
    =\sum_{u,v} P_{UV}(u,v)P_{X|au}=P_{X|a}.
  \end{split}
\end{equation}
This may seem like an innocuous implication; we only want to study
nonsignaling models in any case.  However, clearly a nonlocal model
can be signaling and therefore cannot in general obey
Eq.~(\ref{eq:6}) nor Eq.~(\ref{eq:2}). Thus, the claim in \cite{CR08}
``that identities (2) do not restrict the generality of the hidden
variable model'' is incorrect.


{\em What's wrong with CR's restriction on a nonlocal HV model?---}
Using CR's restriction (\ref{eq:2}) we will find that explicitly local
HVs within a HV model [that obeys (\ref{eq:3})] must belong to the
``nonlocal part'' to mask out the signaling properties of the
``nonlocal part.''

Let $U=V$ be uniformly distributed between $0$ and $2\pi$, $W=A$, and
let the outcome probabilities be given by
\begin{equation}
\label{eq:7}
  \begin{split}
    P_{X|abuvw}(+1)&=P_{X|au}(+1)=
    \begin{cases}
      1,&a<u<a+\pi,\\
      0,&\text{otherwise}.
    \end{cases}\\
    P_{Y|abuvw}(-1)&=P_{Y|bvw}(-1)\\&=
    \begin{cases}
      1,&w+\pi\sin^2(b-w)<v\\
      &\quad<w+\pi\sin^2(b-w)+\pi,\\
      0,&\text{otherwise}.
    \end{cases}\\
  \end{split}
\end{equation}
This gives the quantum predictions from the singlet state (see below).

The nonlocal HV $W$ only carries information about Alice's measurement
setting, and is independent of the local HVs. There can be no doubt
that $U$ and $V$ are local HVs; their distributions do not depend on
$a$, $b$, or the value $w$ of $W$; and the outcomes depend only on the
local setting and the HVs that are available at each site, so that
(\ref{eq:3}) is fulfilled. Indeed, $U$ also fulfils CR's definition of
local part, (\ref{eq:2}). Remarkably, even so, $U$ belongs to CR's
``nonlocal part,'' but to see this we need to take a complicated
route. We first must verify that $V$, in spite of being a local HV as
noted above, belongs to CR's ``nonlocal part'' by taking the average
of $P_{Y|bvw}$ over $W$ (here, using an integral rather than a sum,
and $p_{W|abuv}(w)=\delta(w-a)$),
\begin{equation}
  \label{eq:8}
  \begin{split}
    P_{Y|abuv}(-1)&=\int_w p_{W|abuv}(w)P_{Y|bvw}(-1)dw\\&=
    \begin{cases}
      1,&a+\pi\sin^2\big(\tfrac{b-a}2\big)<v\\
      &\qquad<a+\pi\sin^2\big(\tfrac{b-a}2\big)+\pi,\\
      0,&\text{otherwise}.
    \end{cases}
  \end{split}
\end{equation}
This average depends on $a$, and this is a direct consequence of the
fact that the distribution of $W$ depends on $a$.  This does not
fulfill Eq.~(\ref{eq:2}), so CR's restriction forces the conclusion
that $V$ belongs to the ``nonlocal part'' in spite of having a local
distribution.  And this means that we need to average over the new
``nonlocal part'' $W'=(V,W)$ to find out whether $U$ belongs to the
``local part,''
\begin{equation}
\label{eq:9}
  \begin{split}
       P_{Y|abu}(-1)&=\iint_{vw} p_V(v)p_{W|abuv}(w)P_{Y|bvw}(-1)dwdv\\&=
    \begin{cases}
      1,&a+\pi\sin^2\big(\tfrac{b-a}2\big)<u\\
      &\qquad<a+\pi\sin^2\big(\tfrac{b-a}2\big)+\pi,\\
      0,&\text{otherwise}.
    \end{cases}
  \end{split}
\end{equation}
And this depends on $a$ as well as $u$ [the remote setting and HV]. The
local HVs $U$ and $V$ thus both belong to the ``nonlocal part.''

For completeness, we use the above expressions to determine the
joint outcome probabilities and find, for example,
\begin{equation}
\label{eq:10}
  \begin{split}
    P_{XY|ab}(+1,+1)&=\frac{[a+\pi\sin^2\big(\tfrac{b-a}2\big)]-a}{2\pi}\\&
    =\tfrac12\sin^2\big(\tfrac{b-a}2\big),\\
  \end{split}
\end{equation}
as desired.

The ``local part'' in CR's sense is trivial, not because of lack of
local HVs, but because all of the the local HVs are drawn into the
``nonlocal part'' by CR's extra requirement.


{\em Conclusions.---} Specific classes of nonlocal HV models have been
studied in a number of papers. These classes are all restricted in one
way or another, which for example is simple to see in \cite{EPR92}.
Also, Aspect \cite{Aspect07} and De Zela \cite{DeZela08} have pointed
out that there are nonlocal HV models that are not addressed by
Leggett's inequality. The apparent strength of CR's result was
precisely that it seems to lead to a more general statement: the
possibility of an experimental refutation of any HV model with a
nontrivial local part.

However, we have here shown that CR's division of nonlocal HVs into
``local part'' and ``nonlocal part'' carries an extra nonsignaling
restriction on the ``nonlocal part'' that is not physically nor
mathematically motivated.  This extra requirement makes the claim
``any hidden variable model can only be compatible with quantum
mechanics if its local part is trivial'' \cite{CR08} unfounded.  The
statement should instead be: Any HV model that splits into a ``local
part'' and a \emph{nonsignaling} ``nonlocal part'' can only be
compatible with quantum mechanics if its ``local part'' is
trivial. Furthermore, a trivial ``local part'' does not mean that the
model lacks local HVs, only that all HVs (even local ones) are forced
into to the ``nonlocal part,'' to prevent signaling within the
``nonlocal part.'' We can only conclude that no startling conclusions
can be drawn from the split into ``local'' and ``nonlocal parts''
proposed in \cite{CR08}.

\begin{acknowledgments}
  The authors thank M.\ Bourennane, R.\ Colbeck, R.~Renner, and
  H.\ Weinfurter for discussions. A.C.~acknowledges support from the
  Spanish MEC Project No. FIS2005-07689, and the Junta de
  Andaluc\'{\i}a Excellence Project No.~P06-FQM-02243.
\end{acknowledgments}



\begin{thebibliography}{99}

\bibitem{EPR35}
A. Einstein, B. Podolsky, and N. Rosen,
Phys. Rev. {\bf 47}, 777 (1935).


\bibitem{ADR82}
A. Aspect, J. Dalibard, and G. Roger,
Phys. Rev. Lett. {\bf 49}, 1804 (1982).

\bibitem{WJSWZ98}
G. Weihs {\em et al.},
Phys. Rev. Lett. {\bf 81}, 5039 (1998).

\bibitem{RKMSIMW01}
M. A. Rowe {\em et al.},
Nature (London) {\bf 409}, 791 (2001).

\bibitem{MMMOM08}
D. N. Matsukevich {\em et al.},
Phys. Rev. Lett. {\bf 100}, 150404 (2008).


\bibitem{RWVHKCZW09}
W. Rosenfeld {\em et al.},
Adv. Sci. Lett. {\bf 2}, 469 (2009).


\bibitem{Bell64}
J. S. Bell,
Physics (Long Island City, NY) {\bf 1}, 195 (1964).

\bibitem{CHSH69}
J. F. Clauser, M. A. Horne, A. Shimony, and R. A. Holt,
Phys. Rev. Lett. {\bf 23}, 880 (1969).


\bibitem{KZGKGCBR09}
G. Kirchmair {\em et al.},
Nature (London) {\bf 460}, 494 (2009).


\bibitem{Specker60}
E. Specker,
Dialectica {\bf 14}, 239 (1960).

\bibitem{Bell66}
J. S. Bell,
Rev. Mod. Phys. {\bf 38}, 447 (1966).

\bibitem{KS67}
S. Kochen and E. P. Specker,
J. Math. Mech. {\bf 17}, 59 (1967).

\bibitem{Cabello08}
A. Cabello,
Phys. Rev. Lett. {\bf 101}, 210401 (2008).


\bibitem{GPKBZAZ07}
S. Gr\"oblacher {\em et al.},
Nature (London) {\bf 446}, 871 (2007); {\em ibid.} {\bf 449}, 252
(2007).

\bibitem{BLGKLS07}
C. Branciard {\em et al.},
Phys. Rev. Lett. {\bf 99}, 210407 (2007).

\bibitem{BBGKLLS08}
C. Branciard {\em et al.},
Nat. Phys. {\bf 4}, 681 (2008).

\bibitem{Leggett03}
A. J. Leggett,
Found. Phys. {\bf 33}, 1469 (2003).

\bibitem{EPR92}
 A. C. Elitzur, S. Popescu and D. Rorlich
 Phys. Lett. A {\bf 162}, 25 (1992).

\bibitem{CR08}
R. Colbeck and R. Renner,
Phys. Rev. Lett. {\bf 101}, 050403 (2008).



\bibitem{CRpriv}
 R. Colbeck and R. Renner (private communication).




\bibitem{Aspect07}
A. Aspect,
Nature (London) {\bf 446}, 866 (2007).

\bibitem{DeZela08}
F. De Zela,
J. Phys. A: Math. Theor. {\bf 41}, 505301 (2008).


\end{thebibliography}
\end{document}